\documentclass[12pt,a4paper]{article}
\usepackage{amsmath,amssymb,epsfig}
\newcommand{\beq}{\begin{eqnarray}}
\newcommand{\eeq}{\end{eqnarray}}

\newcommand{\be}{\begin{equation}}
\newcommand{\ee}{\end{equation}}

\newcommand{\bm}{\begin{multline}}
\newcommand{\fm}{\end{multline}}

\begin{document}
\setlength{\unitlength}{.8mm}
\numberwithin{equation}{section}

\begin{titlepage} 
\vspace*{0.5cm}
\begin{center}
{\Large\bf Finite volume spectrum of $N=1$ superminimal models perturbed by $\Phi_{13}$}
\end{center}
\vspace{1.5cm}
\begin{center}
{\large \'Arp\'ad Heged\H us}
\end{center}
\bigskip

\vspace{0.1cm}

\begin{center}
Research Institute for Particle and Nuclear Physics,\\
Hungarian Academy of Sciences,\\
H-1525 Budapest 114, P.O.B. 49, Hungary\\ 
\end{center}
\vspace{2.5cm}
\begin{abstract}

We describe an extension of the nonlinear integral equation (NLIE) tehnique to $N=1$ superminimal models 
perturbed by $\Phi_{13}$. Along the way, we also complete our previous studies of the finite volume 
spectrum of the $N=1$ supersymmetric sine-Gordon model by considering the attractive regime and more 
specifically, breather states.

\end{abstract}

\end{titlepage}

\section{Introduction}

Finite size effects play an important role in the investigation of quantum field theories.
They provide a possibility to determine many important physical characteristics of the models such as
S-matrices, mass ratios, relations between parameters appearing in the ultraviolet (UV) and infrared (IR) 
descriptions of the theory and a great deal of qualitative information on the spectrum. 
The investigation of finite size effects in integrable 1+1 dimensional integrable quantum field theories
is particularly useful due to the wide range of exact techniques applicable.
 
Several exact methods have been worked out to study finite size effects in integrable quantum field theories
\cite{1,1a,2,3}. Among all these methods the so called  nonlinear-integral equation (NLIE)
technique \cite{2,3} proved to be the most efficient method to study finite size effects in 
this family of quantum field theories.
 In most of the cases these equations can be derived from an integrable Bethe Ansatz solvable
lattice regularization of the model. 

With the help of the NLIE technique the best studied model is the sine-Gordon model. The integrable lattice 
regularization starting from which the NLIE can be derived is the inhomogeneous 6-vertex model with 
alternating inhomogeneities \cite{3}. The NLIE is derived for the lattice model, then tuning appropriately
the lattice constant and the inhomogeneity parameter one gets the NLIE for the continuum sine-Gordon model
\cite{4}.
This NLIE was shown to describe the full finite size spectrum of the sine-Gordon model \cite{5a,5b}, and 
moreover
considering twisted versions of the NLIE one can describe finite size effects of massive integrable
perturbations of Virasoro minimal models (unitary and non-unitary) \cite{6}.   

In this paper we investigate the $N=1$ supersymmetric sine-Gordon model (SSG) by means of the NLIE technique.
By analogy with the sine-Gordon model we consider the inhomogeneous 19-vertex model \cite{7} with 
alternating inhomogeneities as integrable lattice regularization of the model. Then using the 
auxiliary functions introduced in \cite{8} and the functional relations among them
provided by the T-Q relations, the NLIE governing the finite size effects of the regularized model can be
obtained \cite{9}. Carrying out the prescribed continuum limit procedure one arrives at the NLIE of the
supersymmetric sine-Gordon model (continuum field theory) \cite{10}. In \cite{10} 
the finite size effects in the repulsive regime of the supersymmetric sine-Gordon model has been studied.

In this paper we complete this earlier analysis by the extension of the NLIE technique to the attractive 
regime of the model as well as to $N=1$ superminimal models perturbed by $\Phi_{13}$.

The paper is organized as follows:
In section 2 after recalling some basic generalities about the supersymmetric sine-Gordon model
we present the twisted NLIE being valid either in the attractive or in the repulsive regime.
Section 3 is devoted to the infrared analysis of the NLIE reconstructing the breather S-matrices.
In section 4 the UV limit of the finite size spectrum predicted by the NLIE is computed 
and shown in the twistless case to agree with the spectrum of the expected modular invariant $c=3/2$ 
conformal field theory (CFT). The relation between the twisted NLIE and the $N=1$ superminimal models
perturbed by $\Phi_{13}$ is discussed in section 5. Section 6 is devoted to discuss some examples of
breather states of the supersymmetric sine-Gordon model. In section 7  we perform some numerical 
checks on our twisted NLIEs. Finally our conclusions and perspectives for future work can be found 
in section 8.

\section{NLIE for the supersymmetric sine-Gordon model in the attractive regime}

\subsection{The model}

The $N=1$ supersymmetric sine-Gordon model (SSG) is defined by the action: 
\begin{equation}
\mathcal{A}_{SSG}=\int dt\int_{0}^{L}dx\left(\frac{1}{2}\partial_{\mu}\varphi\partial^{\mu}\varphi+i\bar{\psi}
\gamma^{\mu}\partial_{\mu}\psi+\mu\psi\bar{\psi}\cos\frac{\beta}{2}\varphi+\frac{\mu^{2}}{\beta^{2}}\cos
\beta\varphi\right),\label{SSG-action}\end{equation}
where $\varphi$ is a real scalar, $\psi$ is a Majoranna fermion field, $\beta$ is the coupling constant
and the dimensionful parameter $\mu$ defines the mass scale, which can be expressed by the kink mass 
$\mathcal{M}$  \cite{11} as follows:
\begin{equation}
\mu=\frac{8 \, \mathcal{M}^{1-\frac{\beta^{2}}{16\pi}}
\left(\frac{\pi}{4}\frac{\beta^{2}}{16\pi-\beta^{2}}\right)^{1-\frac{\beta^{2}}{16\pi}}}
{\gamma\left(\frac{1}{2}-\frac{\beta^{2}}{32\pi}\right)},
\qquad\gamma(x)=\frac{\Gamma(x)}{\Gamma(1-x)}.
\end{equation}
For later convenience, we define a new parameter $p$ by
\begin{equation}
p=\frac{2 \beta^2}{16 \pi-\beta^2}.
\end{equation}
In the repulsive regime $p>1$ while in the attractive $0<p<1$. The $k$-th breather threshold is $p=1/k$.

The UV limit of the theory is a $c=3/2$ CFT being composed of a Neveu-Schwartz (NS)
and a Ramnond (R) sectors corresponding to antiperiodic and periodic boundary conditions imposed on
the Majoranna fermion respestively.  
In the NS sector the primary fields are given by vertex operators
\begin{equation} \label{NSVnm}
V_{n,m}^{(r,\bar{r})}(z,\bar{z})=\bar{\psi}_{\bar{r}}(\bar{z})\psi_{r}(z):e^{i\left[\left(\frac{n}{R}
+\frac{mR}{2}\right)\phi(z)+\left(\frac{n}{R}-\frac{mR}{2}\right)\bar{\phi}(\bar{z})\right]}:,
\end{equation}
where $r,\bar{r} \in \{0,1 \}$, $\psi_0=1$, $\psi_1=\psi$ of conformal dimensions
$$ \Delta_{n,m}^{(r,\bar{r})}=\frac{1}{2}\left(\frac{n}{R}+\frac{mR}{2}\right)^{2}+\frac{r}{2}
\quad, \quad\bar{\Delta}_{n,m}^{(r,\bar{r})}=\frac{1}{2}\left(\frac{n}{R}-\frac{mR}{2}\right)^{2}
+\frac{\bar{r}}{2},$$
while in the R sector the vertex operators corresponding to the primary fields 
have conformal weights 
$$ \Delta_{n,m}=\frac{1}{2}\left(\frac{n}{R}+\frac{mR}{2}\right)^{2}+\frac{1}{16}\quad,\quad
\bar{\Delta}_{n,m}=\frac{1}{2}\left(\frac{n}{R}-\frac{mR}{2}\right)^{2}+\frac{1}{16}$$
and read as
\begin{equation} \label{RVnm}
R_{n,m}(z,\bar{z})=\sigma(z,\bar{z})
:e^{i\left[\left(\frac{n}{R}+\frac{mR}{2}\right)\phi(z)+\left(\frac{n}{R}-\frac{mR}{2}\right)
\bar{\phi}(\bar{z})\right]}:, 
\end{equation}
with $\sigma(z,\bar{z})$ being the Ising spin field.

The allowed set of integer or half-integer valued quantum numbers 
$n,m$ of the above vertex operators (\ref{NSVnm}), (\ref{RVnm}) are encoded into 
the modular invariant partition function as follows:
\begin{eqnarray} \label{ZR}
Z(R) & = & \frac{1}{|\eta(q)|^{2}}\left\{ 
(\chi_{0}(q)\bar{\chi}_{1/2}(q)+\chi_{1/2}(q)\bar{\chi}_{0}(q))\sum_{n\in\mathbb{Z}+\frac{1}{2},\, 
m\in2\mathbb{Z}+1}\right.\\
 & + & \left.(|\chi_{0}(q)|^{2}+|\chi_{1/2}(q)|^{2})\sum_{n\in\mathbb{Z},\, 
m\in2\mathbb{Z}}+|\chi_{1/16}(q)|^{2}\sum_{2n-m\in2\mathbb{Z}+1}\right\} 
q^{\Delta_{n,m}^{+}}\bar{q}^{{\Delta}_{n,m}^{-}} \nonumber
\end{eqnarray}
where 
$$ \Delta_{n,m}^{\pm}=\frac12 \left( \frac{n}{R}\pm \frac{m}{2}R\right)^2, $$
 are the conformal weights of the Gaussian part of the CFT, 
$\eta(q)$ is the Dedekind function, $\{ \chi_{0}(q),\chi_{1/2}(q),\chi_{1/16}(q) \}$ 
are the Virasoro characters corresponding
to the irreducible representations of a $c=1/2$ CFT, $q=e^{2\pi i\tau}$, $\tau$ being the modular parameter.
The compactification radius $R$ related to $p$ as follows:
\begin{equation} \label{comprad}
R=\sqrt{1+\frac{2}{p}}.
\end{equation}

\subsection{Nonlinear integral equations}

We briefly recall the NLIE for the (twisted) SSG. We use the notations and conventions 
of the papers \cite{10} and \cite{12}, which the reader is invited to consult for more details. We put the 
SSG on a cylindrical spacetime, with infinite time direction and compact spatial 
extension of length (volume) $L$. The NLIEs governing the finite size effects of the (twisted) SSG
 read as follows:
\begin{eqnarray}
\log b(\theta) &=& C_b+i D(\theta)+i g_1(\theta)+i g_b(\theta)+(G*_{_{\Gamma}} \ln B)(\theta)-
 (G*_{_{\bar\Gamma}} \ln {\bar B})(\theta) \nonumber \\
&+& \lim_{\varepsilon \to 0^+}(K^{+\frac{\pi}{2}-\varepsilon}*\ln Y)(\theta)  \label{nlie1} \\
\log y(\theta)&=& C_y+i g_y(\theta)+(K^{+\frac{\pi}{2}}*_{_{\Gamma}} \ln B)(\theta)+
 (K^{-\frac{\pi}{2}}*_{_{\bar\Gamma}} \ln {\bar B})(\theta), \nonumber
\end{eqnarray}
where $Y(\theta)=1+y(\theta), \quad B(\theta)=1+b(\theta)$ and $\bar{B}(\theta)$ stands for
the complex conjugate of $B(\theta)$.
We introduced the notation for any function $f$
$$ f^{\pm \eta}(\theta)=f(\theta\pm i \eta).$$
The eqs. (\ref{nlie1}) contain three type of convolutions, one of them is the usual 
one containing integration along the real axis
$$ (f*g)(x)=\int\limits_{-\infty}^{\infty} dy \, f(x-y) g(y), $$ 
while the other two ones are defined by integrating on the complex plane along the 
integration contours $\Gamma(t)$ and $\bar\Gamma(t) \qquad (t \in \mathbb{R})$: 
$$  (f*_{_{\Gamma}} g)(x)=\int_{\Gamma} dz \, f(x-z) g(z), \qquad 
(f*_{_{\bar\Gamma}} g)(x)=\int_{\bar\Gamma} dz \, f(x-z) g(z), $$ 
where the curve $\bar\Gamma(t)$ is the complex conjugate of $\Gamma(t)$. 
The continuous non-self-intersecting contour $\Gamma(t)$ fulfills the following properties:  
\newline
\newline
1. $\mbox{Re}\Gamma(\pm \infty)=\pm \infty$,
\newline
\newline
2. $0\leq\mbox{Im}\Gamma(t)< \mbox{min}(1,p)\, {\pi}/{2} \qquad \forall t \in {\mathbb R} $.
\newline
\newline
The kernel functions $G$ and $K$  of (\ref{nlie1}) read as
  \begin{equation} \label{GK}
 G(\theta)=\int\limits _{-\infty}^{\infty}\frac{dq}{2\pi}\,\, 
e^{iq\theta}\,\,\frac{\sinh\frac{\pi(p-1)q}{2}}{2\sinh\frac{\pi pq}{2}\cosh\frac{\pi q}{2}},\qquad 
K(\theta)=\frac{1}{2\pi\cosh(\theta)}.
\end{equation}
We also introduce the odd primitives of the kernel functions (see appendix D. for the choice of
 branch cuts of $\chi_{K}(\theta)$), 
 \begin{equation} \label{chi}
\chi(\theta)=2\pi\int\limits _{0}^{\theta}dx\,\, G(x),\qquad\chi_{K}(\theta)=2\pi\int\limits 
_{0}^{\theta}dx\,\, K(x), 
\end{equation}
that are important in writing the source terms containing information
on the excitations:
\begin{eqnarray}
g_{b}(\theta) & = & \sum_{j=1}^{N_{H}}\chi(\theta-h_{j})+
\sum_{j=1}^{N_{V}^{S}}\left(\chi(\theta-v_{j})+\chi(\theta-\bar{v}_{j})\right)
-\sum_{j=1}^{N_{S}}\left(\chi(\theta-s_{j})+\chi(\theta-\bar{s}_{j})\right) \nonumber \\
 & - & \sum_{j=1}^{M_{C}}\chi(\theta-c_{j})-\sum_{j=1}^{M_{W}}\chi_{II}(\theta-w_{j})
 -\sum_{j=1}^{M_{sc}}\chi_{II}(\theta-w_{sc}^{(j)}),\\
g_{1}(\theta) & = & \sum_{j=1}^{N_{1}}\chi_{K}(\theta-h_{j}^{(1)}),\\
g_{y}(\theta) & = & \lim_{\eta\rightarrow 0^{+}}\tilde{g}_{y}\left(\theta+i\frac{\pi}{2}-i\eta\right),
\nonumber \\
\tilde{g}_{y}(\theta) & = & 
\sum_{j=1}^{N_{H}}\chi_{K}(\theta-h_{j})
+\sum_{j=1}^{N_{V}^{S}}\left(\chi_{K}(\theta-v_{j})+\chi_{K}(\theta-\bar{v}_{j})\right)
-\sum_{j=1}^{M_{S}}\left(\chi_{K}(\theta-s_{j})+\chi_{K}(\theta-\bar{s}_{j})\right) \nonumber \\
 & - & \sum_{j=1}^{M_{C}}\chi_{K}(\theta-c_{j})
 -\sum_{j=1}^{M_{W}}\chi_{KII}(\theta-w_{j})
 -\sum_{j=1}^{M_{sc}}\chi_{KII}(\theta-w_{sc}^{(j)}), \end{eqnarray}
 where the second determination of any function: $f_{II}(\theta)$ is defined as in \cite{5a}
   \begin{equation}\label{detII}
	f_{II}(\theta)= \begin{cases}
	f(\theta)+f(\theta-i\, \pi\,\mbox{sign}(\mbox{Im} \, \theta))\; & 1<p \\ 
	f(\theta) - f(\theta-i \, p \, \pi \, \mbox{sign}(\mbox{Im} \, \theta)) \; & 0<p<1. 
		   \end{cases}
\end{equation}
The objects appearing in the source terms of (\ref{nlie1}) are as follows \cite{12}:
\newline
1. \emph{Type I holes}: \, $\{h_j^{(1)}\}, \quad j=1,\dots,N_1$
\newline
\newline
2. \emph{Holes}: \, $\{h_j\}, \quad j=1,\dots,N_H$ 
\newline
\newline
3. "\emph{Close source objects}": $\{c_j\}$ $\quad j=1,\dots,M_C$, 
satisfying the condition:
\newline $\mbox{Im} \Gamma(\mbox{Re} \, c_j)<|\mbox{Im} \, c_j|<\mbox{min}(1,p) \, \pi.$
\newline
\newline
4. \emph{Wide effective roots}: \, $\{w_j\}, \quad j=1,\dots,M_W$ 
\newline
\newline
5. \emph{Self-conjugate effective roots}: \, $\{w_{sc}^{(j)}\}, \quad j=1,\dots,M_{sc}$
\newline
\newline
6. \emph{Ordinary special objects}: \, $\{s_j\}, \, \mbox{and their complex conjugates} \, \{\bar{s}_j\}
 \quad j=1,\dots,N_S$ 
 \newline
 \newline
7. \emph{Virtual special objects}: \, $\{v_j\}, \, \mbox{and their complex conjugates} \, \{\bar{v}_j\}
 \quad j=1,\dots,N_V^S$.
\newline
\newline
Using the light-cone lattice approach \cite{3} the NLIE (\ref{nlie1}) was derived from the inhomogeneous
19-vertex model with alternating inhomogeneities. Thus the source objects of (\ref{nlie1}) are related to
the Bethe roots and the zeroes of transfer matrices of the fusion hierarchy of the underlying 19-vertex 
model. This relation and the detailed description of ordinary and virtual special objects can be found
 in \cite{12}.
The driving term bulk contribution in the equation for $\ln b(\theta)$ reads as 
 $$D(\theta)=\ell \, \sinh\theta,$$
 where the dimensionless scale parameter ${\ell}={\mathcal M} L$ with ${\mathcal M}$ and $L$
 being the kink mass and the volume respectively.  
 
 The values of the constants of the NLIE (\ref{nlie1}) are as follows
 $$ C_b=i (\pi \, \delta_b+\alpha), \quad 
 \alpha=\omega\left(1+\frac{2}{p}\right)+\chi_{\infty}(N_{-}-N_{+}), 
  \qquad \delta_b \in \{0,1 \},$$
$$ N_{\pm}=\left[ \frac{3S}{p+2}\mp \frac{3\omega}{\pi}\right]-\left[ \frac{S}{p+2}\mp 
\frac{\omega}{\pi}\right], \qquad \chi_{\infty}=\chi(+\infty)=\frac{\pi}{2}\left(1-\frac{1}{p} \right),$$
and
$$ C_y=i \, \pi \,(S+\delta_y)+ i \, \pi \, \Theta(p-1) \,(M_W+M_{sc}), \qquad \delta_y \in \{0,1 \},$$ 
where $[...]$ stands for integer part, $\Theta(x)$ denotes the Heaviside function and $\omega$
is the twist angle of the underlying solvable lattice model. For conventions used here see 
section 3 of ref. \cite{10}. 
In addition to (\ref{nlie1}) we need two other equations for the determination of type I holes and 
effective wide and self-conjugate roots.

The positions of type I holes can be determined from the function $a(\theta)$ given by:
\begin{equation} \label{lna1}
-\log a(\theta)=i \, \delta_a(\theta)+(K*_{_{\Gamma}} \ln B)(\theta)-
 (K*_{_{\bar\Gamma}} \ln {\bar B})(\theta)-C_y, \qquad 0\leq |\mbox{Im} \, \theta|< \frac{\pi}{2},
\end{equation} 
\begin{eqnarray}
\delta_a(\theta) & = & 
\sum_{j=1}^{N_{H}}\chi_{K}(\theta-h_{j})
+\sum_{j=1}^{N_{V}^{S}}\left(\chi_{K}(\theta-v_{j})+\chi_{K}(\theta-\bar{v}_{j})\right)
-\sum_{j=1}^{M_{S}}\left(\chi_{K}(\theta-s_{j})+\chi_{K}(\theta-\bar{s}_{j})\right) \nonumber \\
 & - & \sum_{j=1}^{M_{C}}\chi_{K}(\theta-c_{j})
 -\sum_{j=1}^{M_{W}}\chi_{KII}(\theta-w_{j}) -\sum_{j=1}^{M_{sc}}\chi_{KII}(\theta-w_{sc}^{(j)}).
  \end{eqnarray}
The function necessary to know for the determination of wide and self-conjugate effective roots 
is as follows, for $\mbox{min}(1,p) \, \pi<\mbox{Im} \, \theta \leq \frac{\pi(p+1)}{2}$:
\begin{eqnarray} 
\log \tilde a(\theta) &=& i \, D_{II}(\theta)+
i \, g_{1II}(\theta) +i \, g_{bII}(\theta)+(G_{II}*_{_{\Gamma}} \ln B)(\theta)-(G_{II}*_{_{\bar\Gamma}} \ln 
{\bar B})(\theta) \label{lnatilde} \nonumber \\ 
&+& ((K^{-\frac{\pi}{2}})_{II}*\ln Y)(\theta)+C_{\tilde{a}},  \label{atilde0}
\end{eqnarray}
where
\begin{equation}\label{detII}
	C_{\tilde{a}}= \begin{cases}
	2 \, i \,  \alpha +i \pi (N_{-}-N_{+})\; & 1<p, \\ 
	i \pi (N_{-}-N_{+}) \; & 0<p<1. 
		   \end{cases}
\end{equation}
The source objects appearing in the NLIE are not arbitrary parameters, but they have to satisfy
certain quantization conditions \cite{9,10,12}.  These are as follows:
\begin{itemize}
\item For holes: \begin{equation}
\frac{1}{i}\,\log b(h_{j})=2\pi\, I_{h_{j}}, \qquad j=1,...,N_{H}.\label{eq:holes}\end{equation}

\item For ordinary special objects: \begin{equation}
\mbox{Im} \log b(s_j)=2 \pi I_{s_j}, \quad |b(s_j)|>1, \quad 
(\mbox{Im} \log b)'(s_j)<0,
\quad j=1,\dots,N_S.\label{eq:specials}\end{equation}

\item For virtual special objects: \begin{equation}
\mbox{Im} \log b(v_j)=2 \pi I_{v_j}, \quad |b(v_j)|>1, \quad 
(\mbox{Im} \log b)'(v_j)>1,
\quad j=1,\dots,N_V^S.\label{eq:virtspecials}\end{equation}
 
\item For close source objects (only for the upper part of the close effective pair): \begin{equation}
\frac{1}{i}\,\log b(c_{j}^{\uparrow})=2\pi\, I_{c_{j}^{\uparrow}}, \qquad 
j=1,...,M_{C}/2.\label{eq:close}\end{equation}
 
\item For wide effective roots: \begin{equation}
\frac{1}{i}\,\log \tilde{a}(w_{j}^{\uparrow})=2\pi\, I_{w_{j}^{\uparrow}}, \qquad 
j=1,...,M_{W}/2.\label{eq:wide}\end{equation}
 
\item For self-conjugate effective roots: \begin{equation}
\frac{1}{i}\,\log \tilde{a}(w_{sc}^{\uparrow(j)})=2\pi\, I_{w_{sc}^{\uparrow(j)}}, \qquad 
j=1,...,M_{sc}.\label{eq:self}\end{equation}
 
\end{itemize}
So far we have determined only the upper part of the complex pairs,
but the other parts can be determined by simple complex conjugation. 

\begin{itemize}
\item Finally for type I holes: \begin{equation}
\frac{1}{i}\,\log a (h_{j}^{(1)})=2\pi\, I_{h_{j}^{(1)}}, \qquad j=1,...,N_{1}.\label{eq:t1}\end{equation}

\end{itemize}
All the above quantum numbers $I_{\alpha_{j}}$s are half integers.
A state is then identified by a choice of the quantum numbers $(I_{h_{j}},I_{c_{j}},...)$.
We also mention that the NLIE itself can impose constraints on the allowed values of some
of these quantum numbers.

The counting equations in the continuum theory read as:
\begin{equation} \label{ce1}
N_H+2N_V^S-2N_S=2S+M_C+2 \Theta(p-1)(M_W+M_{sc}),
\end{equation}
\begin{equation} \label{ce2}
N_1-2N_R^S=S+M_1-M_R-\delta_y, 
\end{equation}
where $S$ is the topological charge of the theory\footnote{In this convention charged kinks carry
topological charge $\pm \frac12$. In our case $S$ can take only integer values because in finite volume
the periodic periodic boundary conditions restrict the number of kinks to be even.}
, $M_1$ denotes the number of effective roots 
(or source objects) with the property $\frac{\pi}{2}<|\mbox{Im} \, \theta| \leq \frac{\pi(p+2)}{2}$, 
$M_R$ stands for the number of those real zeroes of the function $1+a(\theta)$ which are not type I
holes. In the lattice Bethe Ansatz they correspond to real Bethe roots. Finally $N_R^S$ denotes the
number of real special objects. They are such real objects where $a(\theta)$ of (\ref{lna1}) takes the value 
$-1$, and $i \,\frac{d}{d \theta} \ln a(\theta)$ is negative.  

In the repulsive regime, the complex roots describe the internal degrees of freedom of kinks.
In the attractive regime, however, configurations consisting entirely of wide and self-conjugate
effective roots describe the breathers and it is clear from (\ref{ce1}) that they do not contribute to the
topological charge.

The energy and momentum of a state can be expressed by the solution of the NLIE as:
\begin{eqnarray}
E & = & \mathcal{M}\left(\sum_{j=1}^{N_{H}}\,\cosh(h_{j})+
\sum_{j=1}^{N_V^{S}}\left\{ \cosh(v_{j})+\cosh(\bar{v}_{j})\right\}
-\sum_{j=1}^{N_{S}}\left\{ \cosh(s_{j})+\cosh(\bar{s}_{j})\right\} \right. \nonumber \\
& - & \sum_{j=1}^{M_{C}}\,\cosh(c_{j})-\sum_{j=1}^{M_{W}}\,\cosh(w_{j})_{II}
-\sum_{j=1}^{M_{W}}\,\cosh(w_{sc}^{(j)})_{II} \label{energia} \\
& + &  \frac{i}{2\pi}\int_{\Gamma} d\theta\sinh(\theta)\,\ln B(\theta)-\frac{i}{2\pi}
\int _{\bar{\Gamma}}d\theta\sinh(\theta)\,\ln\bar{B}(\theta) \Bigg) \nonumber 
\end{eqnarray}
\begin{eqnarray}
P & = & \mathcal{M}\left(\sum_{j=1}^{N_{H}}\,\sinh(h_{j})+-\sum_{j=1}^{N_V^{S}}\,\left\{ 
\sinh(s_{j})+\sinh(\bar{s}_{j})\right\}
-\sum_{j=1}^{M_{S}}\,\left\{ \sinh(s_{j})+\sinh(\bar{s_{j}})\right\} \right. \nonumber \\
 & - & \sum_{j=1}^{M_{C}}\,\sinh(c_{j})-\sum_{j=1}^{M_{W}}\,\sinh(w_{j})_{II}
-\sum_{j=1}^{M_{W}}\,\sinh(w_{sc}^{(j)})_{II} \label{impulzus} \\
  & + &  \frac{i}{2\pi}\int_{\Gamma} d\theta\cosh(\theta)\,\ln 
B(\theta)-\frac{i}{2\pi}\int_{\bar{\Gamma}} d\theta\cosh(\theta)\,
\ln\bar{B}(\theta)\Bigg)\nonumber 
\end{eqnarray}
We notice that there is no bulk energy term due to the supersymmetry of the model.

\section{Infrared limit and breather S-matrices}

In the infrared limit $(L \to \infty)$ the bulk driving term $D(\theta)$ develops a large imaginary part
giving restrictions on the imaginary parts of the positions of close source objects in the infrared limit. 
In the repulsive regime in (\ref{atilde0}) $D_{II}(\theta)=0$ thus there are no restrictions on the
positions of the wide effective roots, 
only the close source objects are forced to form quartets or 2-strings \cite{12}.
 However in the attractive regime $D_{II}(\theta)$ in (\ref{atilde0}) is no longer zero imposing constraints
on the imaginary parts of the wide effective roots as well.  
Analyzing the NLIE and following the argumentation of \cite{5a}, the effective 
roots fall into the configurations as follows:
\newline
\newline
1. \emph{Arrays of the first kind} are effective root configurations containing close source objects and wide 
effective roots as well. These type of arrays consist of effective roots of the form:
$$ \tilde{\theta_k}^{(1)\pm}=\theta \pm i \left(\mu-k p \pi \right), \quad 
\tilde{\theta_k}^{(2)\pm}=\theta \pm i \left( \pi-\mu-(k-1)p \,\pi \right),
\quad k=0,\dots,\left[\frac{1}{2p} \right],$$
with $\theta$ and $0<\mu$ being real parameters. At certain special values of $\mu$ these arrays
degenerate. 
There are two degenerate cases: \emph{odd degenerate arrays}, which contain a self-conjugate effective root at 
$$ \tilde\theta_{sc}=\theta+i \frac{\pi(p+1)}{2}, \qquad \quad \theta \in \mathbb{R}$$ and
accompanying complex pairs at
$$ \tilde\theta_k=\theta \pm i \frac{\pi (1-(2k+1)p)}{2}, \quad k=0,\dots,\left[\frac{1}{2p} \right],$$
and \emph{even degenerate ones}, which contain complex-pairs of effective roots at the positions
$$ \tilde\theta_k=\theta \pm i \frac{\pi (1-2 k p)}{2}, \quad k=0,\dots,\left[\frac{1}{2p} \right].$$
These degenerate arrays contain exactly one pair of close source objects. According to the counting equation 
(\ref{ce1}) the presence of close source objects entails the appearance of holes which correspond to the 
rapidities of the kinks. Thus these arrays of the first kind describe multiparticle configurations 
containing kinks and antikinks. 
\newline
\newline
2. \emph{Arrays of the second kind} contain only wide- and self-conjugate effective roots. They describe the
breather degrees of freedom of the model. There are two types of them.
The \emph{odd ones} contain a self-conjugate effective root 
$$ \tilde\theta_{sc}=\theta+i \frac{\pi(p+1)}{2}, \qquad \quad \theta \in \mathbb{R}$$ and wide effective-pairs 
at
$$ \tilde\theta_k=\theta \pm i \frac{\pi (1-(2k+1)p)}{2}, \quad k=0,\dots,s, \quad
0\leq s \leq \left[\frac{1}{2p} \right]-1,$$
while the \emph{even ones} contain only wide effective-pairs
$$ \tilde\theta_k=\theta \pm i \frac{\pi (1-2 k p)}{2}, \quad k=0,\dots,s, \quad
0\leq s \leq \left[\frac{1}{2p} \right]-1.$$
They correspond to the $(2s+1)$-th breather $B_{2s+1}$ and the $(2s+2)$-th breather $B_{2s+2}$, respectively.

The deviations of the imaginary parts of the previous effective root configurations from the 
formulae listed above are exponentially small in the volume $L$. 

It can be seen that arrays of the second kind become degenerate ones of the first kind, if one analytically 
continues increasing $p$. The reason is that breathers are of course kink-antikink bound states, while
degenerate arrays of the first kind describe scattering states of a kink and an antikink.

In the infrared limit one can drop all terms containing the convolution of $\ln B(\theta)$ and 
$\ln \bar{B}(\theta)$, because they are exponentially small in the volume $L$. One can therefore compute
the energy and momentum contribution of an array of the second kind corresponding to the breather $B_s$.
The energy-momentum contribution turns out to be
$$ (E, \,P)=2 {\cal M} \sin\left(\frac{\pi s p}{2}\right) \, (\cos \theta, \, \sin \theta),$$
where $\theta$ is the common real part of the roots composing the array. This is just the contribution
of a breather $B_s$ moving with rapidity $\theta$.
Arrays of the first kind do not contribute to the energy and momentum in the infrared limit lending support
to their interpretation as polarization states of the kinks. 

Now we proceed to show that with the above interpretation the NLIE correctly reproduces the 2-body scattering
matrices of the SSG including breathers. In the repulsive regime the 2-kink S-matrices
have been calculated in detail in \cite{10}.

Hereafter we will concentrate only on breather-breather 2-body scatterings, because from 
the NLIE or equivalently from the study of the finite size effects of our model
 the direct determination of kink-breather 2-particle scattering amplitudes is not possible.
 This is because according to the counting equation (\ref{ce1}) only even number of kinks and antikinks
 are allowed in finite volume. So the simplest scattering amplitude containing breathers and kinks too,
  which can be extracted from the finite volume analysis of the model contains 2-kinks and one breather, which
  is a 3-body scattering amplitude.  

 \subsection{Breather S-matrices}
 
 For the sake of simplicity we will show that the NLIE correctly reproduces all the 4 eigenvalues of the
 2-body scattering matrix of the first breathers. 
In the language of the NLIE the first breathers are represented simply by self-conjugate effective roots. 
Let the two self-conjugate roots corresponding to the
two $B_1$ breathers:
$$w_{sc}^{(1)}=\theta_1+i \frac{\pi(p+1)}{2}, \quad \mbox{and} \quad w_{sc}^{(2)}=\theta_1+i 
\frac{\pi(p+1)}{2}, \qquad \theta_1, \theta_2 \in \mathbb{R}.$$
Then the S-matrix eigenvalues can be read off from the quantization conditions of $\theta_1$ and 
$\theta_2$ given by the auxiliary function $\log \tilde{a}(\theta)$. 
In the infrared limit, as we have already mentioned before, all terms containing the convolution of 
$\ln B(\theta)$ or $\ln \bar{B}(\theta)$ can be dropped, nevertheless the convolutions containing
$\ln Y(\theta)$ must be kept, because $Y(\theta)$ tends to a finite non-trivial function in the
infrared limit. So in the infrared limit the quantization condition (\ref{eq:self}) reads as
\begin{eqnarray} \label{qat1}
\log \tilde{a}(w_{sc}^{(1)}) &=& i \ell  \sinh_{II}(w_{sc}^{(1)})+I_h^{(0)}(w_{sc}^{(1)})
-i(\chi_{II}^{+\pi/2})_{II}(w_{sc}^{(1)}-w_{sc}^{(2)}-i\frac\pi2) \nonumber \\
&+&
i \sum\limits_{\{ h_j^{(1)}\}} \chi_{KII}(w_{sc}^{(1)}-h_j^{(1)}) 
- i (\chi_{II}^{+\pi/2})_{II}(-i\frac\pi2), \label{qwsc1}
\end{eqnarray}
where 
\begin{equation} \label{Ih0}
I_h^{(0)}(w_{sc}^{(1)})=\int\limits_{-\infty}^{\infty} dy \, (K^{-\pi/2})_{II}(w_{sc}^{(1)}-y) \, \ln 
Y(\theta).
\end{equation}
The bulk source term of (\ref{qwsc1}) gives the phase shift of the two breather wave function:
$$i \ell  \sinh_{II}(w_{sc}^{(1)}) \rightarrow -i \, 2 {\cal M} \sin\frac{p \pi}{2} \, L \, \sinh \theta_1, 
$$
while the third term on the right hand side of (\ref{qwsc1}) always provides the sine-Gordon part of the
expected S-matrix eigenvalue:
$$ -i(\chi_{II}^{+\pi/2})_{II}(w_{sc}^{(1)}-w_{sc}^{(2)}-i\frac\pi2) \rightarrow
-i \, \ln S_{SG}^{(1,1)}(\theta_1-\theta_2).$$
(See appendix A. for the SSG breather-breather S-matrix amplitudes.)
The other terms in (\ref{qat1}) describe the supersymmetry factors of the breather S-matrices.
The four eigenvalues of the $B_1-B_1$ S-matrix correspond to states with
different values of $\delta_y$ and different number of type I holes.
According to the counting equation (\ref{ce2}) in case of $\delta_y=0$ (Neveu-Schwartz sector) 
the value of $N_1$ can be either zero or 2, while in case of $\delta_y=1$ (Ramond sector)
$N_1$ is either zero or 1. 
\newline
{\bf The $\delta_y=1$ case:}
\newline
In this case in the infinite volume limit the function $Y(\theta)$ reads as
\begin{equation} \label{y11} 
Y^{(\delta_y=1)}_{\infty}(\theta)=
\frac{\sin \left(\frac{p \pi }{2}\right) \, \cosh \left( \frac{\theta_1-\theta_2}{2}\right) 
\cosh(\theta-\theta_{12})}
{   \prod\limits_{j=1}^{2} \sinh\left(   \frac{i \pi}{4}+ \frac{i p \pi}{4}-\frac{\theta-\theta_j}{2}
\right) \sinh\left(   \frac{i \pi}{4}+ \frac{i p \pi}{4}+\frac{\theta-\theta_j}{2}
\right)}, \qquad \theta_{12}=\frac{\theta_1-\theta_2}{2}.                     
\end{equation}
Then substituting this into the integral $I_h^{(0)}(w_{sc})$ (\ref{Ih0}) one gets:
\begin{eqnarray}
I_h^{(0)}(w_{sc})|_{w_{sc}=\theta+i \frac{\pi(p+1)}{2}} &=& 
\ln \frac{\cosh\left( \frac{\theta-\theta_1-i p \pi}{4}  
\right) }{\cosh\left( \frac{\theta-\theta_1+i p \pi}{4} \right)}+
\ln \frac{\cosh\left( \frac{\theta-\theta_2-i p \pi}{4}  
\right) }{\cosh\left( \frac{\theta-\theta_2+i p \pi}{4} \right)} 
-\ln \frac{\cosh\left( \frac{\theta-\theta_{12}-i p \pi}{4}  
\right) }{\cosh\left( \frac{\theta-\theta_{12}+i p \pi}{4} \right)} \nonumber \\
&-& i \, \chi_G(\theta-\theta_1)-i \, \chi_G(\theta-\theta_2), \label{int1}
\end{eqnarray}
where 
\begin{equation} \label{chiG}
\chi_G(\theta)=\int\limits_0^{\infty}  \frac{dq}{q} \, \sin(\theta q) \, 
\frac{ \sinh\left(\frac{\pi p}{2} q \right) \, \sinh\left( \left(1-\frac{p}{2} \right)\pi q \right) }
{ \cosh^2\left(\frac{\pi q}{2}\right) \, \cosh(\pi q)  }.
\end{equation}
Then inserting this result into (\ref{qat1}) and taking its exponential in the $N_1=0$ case one gets
the eigenvalue $\Lambda_{+}^{(1)}(\theta_1-\theta_2)$ of (\ref{ev1pm}) of appendix A.
In the $N_1=1$ case the position of the single type I hole in the infrared limit is $h_1^{(1)}=\theta_{12}.$
Then putting everything together the quantization condition (\ref{qat1}) gives the eigenvalue 
$\Lambda_{-}^{(1)}(\theta_1-\theta_2)$ of (\ref{ev1pm}).
\newline
\newline
{\bf The $\delta_y=0$ case:}
\newline
In this case in the infinite volume limit the function $Y(\theta)$ reads as
\begin{equation} \label{y12} 
Y^{(\delta_y=0)}_{\infty}(\theta)=
\frac{\cosh\left( 2(\theta-\theta_{12})\right)+\cosh(\theta_1-\theta_2)+2 \sin^2\left(\frac{\pi p}{2}\right)}
{ 4 \,  \prod\limits_{j=1}^{2} \sinh\left(   \frac{i \pi}{4}+ \frac{i p \pi}{4}-\frac{\theta-\theta_j}{2}
\right) \sinh\left(   \frac{i \pi}{4}+ \frac{i p \pi}{4}+\frac{\theta-\theta_j}{2}
\right)}, \qquad \theta_{12}=\frac{\theta_1-\theta_2}{2}.                     
\end{equation}
Then substituting this into the integral $I_h^{(0)}(w_{sc})$ (\ref{Ih0}) one gets the result summarized in
appendix B. Then inserting the result of this integral into the quantization equation (\ref{qat1}) and putting 
everything together one gets that in case of $N_1=0$ the S-matrix eigenvalue 
$\Lambda_{+}^{(2)}(\theta_1-\theta_2)$ of (\ref{ev2pm}) is reproduced correctly, while 
the other eigenvalue $\Lambda_{-}^{(2)}(\theta_1-\theta_2)$ of (\ref{ev2pm}) is obtained from the
 $N_1=2$ case where the positions of the type I holes are $h^{(1)}_\pm=\theta_{12}+\ln A_{\pm}$ with  
 $A_{\pm}$ is given by (\ref{A+-}).

Thus we demonstrated that all the 4 eigenvalues of the $B_1-B_1$ S-matrix can be reproduced by the NLIEs 
(\ref{nlie1})-(\ref{eq:t1}). 
All the 4 two $B_1$ states are described by the same effective root configuration, 
but they differ in the number of type I holes and in the choice for $\delta_y$. 
 To close this section we also mention that the breather S-matrices can be determined solely from 
the function  $\log \tilde{a}(\theta)$, so the breather S-matrices calculated from the NLIEs are insensible
 for the value of the constant $\delta_b$ and $\alpha$ of (\ref{nlie1}). However, as we will see in the
 next section, these parameters strongly influence the UV conformal weights of the states.

\section{UV limit and kink approximation\label{sec:UV}}

In this section in the presence of a twist angle $\omega$ we examine the UV limit of the states
 described by the NLIEs (\ref{nlie1})-(\ref{eq:t1}). 
 We follow the standard approach described in detail in \cite{2,5a,4Fr,20Fr}.
The positions of the sources for $\ell\rightarrow0$ can remain finite
(central objects), or they can move towards the two infinities as
$\pm\ln\left(\frac{2}{\ell}\right)$ (left\emph{/}right movers). We
introduce the finite parts $\theta_{j}^{\pm,0}$ of their positions
$\theta_{j}$ by subtracting the divergent contribution: \[
\{\theta_{j}\}\rightarrow\left\{ \theta_{j}^{\pm}\pm\ln\left(\frac{2}{\ell}\right),\,\theta_{j}^{0}\right\} .\]
 We denote the number of right/left moving and central objects by
$N_{H}^{\pm,0},N_{S}^{\pm,0},M_{C}^{\pm,0},\dots$ etc. For later
convenience we introduce the right/left moving and central spin given
by \begin{equation}
S^{\pm,0}=\frac{1}{2}(N_{H}^{\pm,0}+2N_V^{S \pm,0}-2N_{S}^{\pm,0}-M_{C}^{\pm,0}-2 \, 
\Theta(p-1)(M_{W}^{\pm,0}+M_{sc}^{\pm,0})).
\end{equation}
 In the UV limit the NLIE splits into three separate equations corresponding
to the three asymptotic regions. This is why for all the auxiliary
functions of the NLIE (\ref{nlie1}) we define the so called
\emph{kink functions} as 
\begin{equation}
F_{\pm}(\theta)=\lim_{\ell\rightarrow0}F\left(\theta\pm\ln\frac{2}{\ell}\right).\qquad F\in\{\log b,\,\log 
y,\log\tilde{a},\dots\}.
\end{equation}
 In the UV limit these kink functions satisfy the so called \emph{kink
equations} which can be straightforwardly determined from (\ref{nlie1}).  
The energy and the momentum in the conformal limit can be expressed 
by the left and right kink functions. 

Following the method worked out in \cite{2,5a,20Fr} it turns out
that in the UV limit the energy and momentum can be expressed by a
sum of dilogarithm functions with the $\theta\rightarrow\pm\infty$
limiting values of the kink functions in their argument. One group of these
two limiting values is trivial, namely: \begin{equation}
b_{\pm}(\pm\infty)=\bar{b}_{\pm}(\pm\infty)=0,\end{equation}
\begin{equation}
y_{\pm}(\pm\infty)=(-1)^{\delta_{y}}.\end{equation}
 The other limiting values satisfy a nontrivial coupled nonlinear
algebraic equations, the so-called plateau equations \cite{5a}:
\begin{eqnarray}
\log b_{\pm}(\mp\infty) & = & {C}_{b}\pm2i\chi_{\infty}(S-2S^{\pm})\pm2\pi i\,\hat{l}_{W}^{\pm}\pm 
i\pi\left(M_{sc}+\frac{N_{1}}{2}-N_{1}^{\pm}\right)\nonumber \\
 & + & \frac{\chi_{\infty}}{\pi}\,\left\{ \ln B_{\pm}(\mp\infty)-\ln\bar{B}_{\pm}(\mp\infty)\right\} 
+\frac{1}{2}\ln Y_{\pm}(\mp\infty),\label{logbplat}\end{eqnarray}
\begin{equation}
y_{\pm}(\mp\infty)=(-1)^{2S^{\pm}+\delta_{y}}\,\left(B_{\pm}(\mp\infty)
\bar{B}_{\pm}(\mp\infty)\right)^{\frac{1}{2}},\label{Nrp}\end{equation}
 where $\hat{l}_{W}^{\pm}$ are integers depending on the relative positions of the complex effective
 roots and their actual value is irrelevant from the point of view of the exponent of equation 
(\ref{logbplat}).
 
 To solve eqs. (\ref{logbplat},\ref{Nrp}) analytically we take the Ansatz of ref. \cite{10}:
\begin{equation}
b_{\pm}(\mp\infty)=e^{\pm3i\rho_{\pm}}\,2\,\cos(\rho_{\pm}),\qquad\bar{b}_{\pm}(\mp\infty)
=e^{\mp3i\rho_{\pm}}\,2\,\cos(\rho_{\pm}),\label{ap1}\end{equation}
\begin{equation}
B_{\pm}(\mp\infty)=e^{\pm2i\rho_{\pm}}\,\frac{\sin\left(3\rho_{\pm}\right)}
{\sin\left(\rho_{\pm}\right)},\qquad\bar{B}_{\pm}(\mp\infty)=e^{\mp2i\rho_{\pm}}\,\frac{\sin\left(3\rho
_{\pm}\right)}{\sin\left(\rho_{\pm}\right)},\label{ap2}\end{equation}
\begin{equation}
y_{\pm}(\mp\infty)=\frac{\sin\left(3\rho_{\pm}\right)}{\sin\left(\rho_{\pm}\right)},\qquad
 Y_{\pm}(\mp\infty)=4\,{\cos(\rho_{\pm})}^{2}>0.\label{ap3}\end{equation}
 Since we need only $b_{\pm}(\mp\infty)$ and not its logarithm, 
 we can restrict the allowed values of $\rho_{\pm}$ in the
$[0,2\pi]$ interval. The solutions of the plateau equations formally
take the form 
\begin{equation}
\rho_{\pm}=\pi\left(k_{\rho}^{\pm}\pm\delta_{b}\pm\frac{\alpha}{\pi}+\Delta_{\rho_{\pm}}\right)-\frac{\pi}{p+2}
\left(2k_{\rho}^{\pm}\pm2\delta_{b}\pm2\frac{\alpha}{\pi}+3\Delta_{\rho_{\pm}}\right),\label{rho1}
\end{equation}
 where 
 \begin{equation}
\Delta_{\rho_{\pm}}=S-2S^{+}-N_{\rho}^{\pm},\qquad N_{\rho}^{\pm}=\left[ 3\frac{\rho_{\pm}}{\pi}
\right] -\left[ \frac{\rho_{\pm}}{\pi}\right] ,\qquad n_{\rho}^{\pm}=\left[ 2\frac{\rho_{\pm}}{\pi}
\right] -\left[ \frac{\rho_{\pm}}{\pi}\right] .
\end{equation}
 and further constraints must be satisfied by the parity of the integers
$k_{\rho}^{\pm}$ and $N_{\rho}^{\pm}$: 
\begin{equation}
k_{\rho}^{\pm}=2l_{\rho}^{\pm}+M_{sc}+\frac{N_{1}}{2}-N_{1}^{\pm}-n_{\rho}^{\pm},\qquad 
l_{\rho}^{\pm}\in\mathbb{Z},\label{cons+}
\end{equation}
\begin{equation}
N_{\rho}^{\pm}\,\,\mbox{mod}\,\,2=2S^{\pm}+\delta_{y}\,\,\mbox{mod}\,\,2\label{N-Sp}
\end{equation}
 Using the dilogarithm sum rule of appendix C and putting everything
together the conformal weights take the form 
\begin{equation}
\Delta^{\pm}=\frac{1}{16}\,\delta_{y}+\frac{1}{2}\left(\frac{n_{\pm}-\frac{\alpha}{\pi}}{R}\pm
\frac{S}{2}R\right)^{2}+\tilde{N}_{\pm}+J_{\pm},\label{DELTA}
\end{equation}
 where 
 \begin{equation}
n_{\pm}=-\left(\delta_{b}\pm k_{\rho}^{\pm}\pm\frac{3}{2}\Delta_{\rho_{\pm}}\right),\label{nplusminus}
\end{equation}
\begin{equation}
\tilde{N}_{\pm}=\frac{\hat{N}_{\rho}^{\pm}-\delta_{y}}{8}\mp 
S^{\pm}n_{\pm}-\frac{3}{2}(S-S^{\pm})S^{\pm},\qquad\hat{N}_{\rho}^{\pm}=N_{\rho}^{\pm}\,\,\mbox{mod}\,\,2.
\label{Npm}
\end{equation}
\begin{eqnarray}
J_{\pm} & = & \pm I_{h^{\pm}}\pm 2I_{v^{\pm}}\mp2I_{s^{\pm}}\mp2I_{c^{\pm\uparrow}}
\mp2I_{w^{\pm\uparrow}}\mp I_{w_{sc}^{\pm\uparrow}}\pm I_{h^{(1)\pm}}\nonumber \\
& \mp & (S^{\pm}+\Theta(p-1)(M_{W}^{\pm}+M_{sc}^{\pm}))\,\left(\delta_{b}\pm\frac{N_{1}}{2}\mp N_{1}^{\pm}\pm 
M_{sc}\mp M_{sc}^{\pm}\right) \label{jpm} \\
 & \mp & \left(\frac{M_{sc}^{\pm}}{2}\right)^2 \mp \frac{N_1^{\pm}}{2} \delta_y
  \mp(M_{W}^{+}+M_{sc}^{+})\,\left(\frac{N_--N_+}{2}\pm(S-S^{\pm}) \right)+K_{\pm},\nonumber 
 \end{eqnarray}
where 
$$I_{h^{\pm}}=\sum\limits_{j=1}^{N_H^\pm} I_{h_j^\pm}, \quad
 I_{s^{\pm}}=\sum\limits_{j=1}^{N_S^\pm} I_{s_j^\pm}, \quad
 I_{v^{\pm}}=\sum\limits_{j=1}^{N_V^{S\pm}} I_{v_j^\pm}, \quad
I_{c^{\pm\uparrow}}=\sum\limits_{j=1}^{M_C^{\pm\uparrow}}  I_{c_j^{\pm\uparrow}},\dots 
\mbox{etc.},$$
and $K_{\pm}$ are integers depending on the concrete complex effective root configuration under consideration.
Analysing the formulae above it can be proven that there is a
relation between $n_{\pm}$ of (\ref{nplusminus}) and the topological charge $S$, 
namely in the $\delta_y=0$ case:
 \begin{equation}
n_{\pm}\in\mathbb{Z}\qquad\mbox{if}\qquad S\in2\mathbb{Z},\label{f1ns}
\end{equation}
\begin{equation}
n_{\pm}\in\mathbb{Z}+\frac{1}{2}\qquad\mbox{if}\qquad S\in2\mathbb{Z}+1,\label{f2ns}
\end{equation}
 while in the  $\delta_y=1$ case:
 \begin{equation}
n_{\pm}\in\mathbb{Z}+\frac{1}{2}\qquad\mbox{if}\qquad S\in2\mathbb{Z},\label{f1r}
\end{equation}
\begin{equation}
n_{\pm}\in\mathbb{Z}\qquad\mbox{if}\qquad S\in2\mathbb{Z}+1.\label{f2r}
\end{equation}
 Moreover it can be proven that depending on the state under
consideration the sum $N_{\pm}+J_{\pm}$ can be either integer or
half-integer, but in the $\delta_{y}=1$ case the sum $N_{\pm}+J_{\pm}$ is always an integer. 

Thus, just like in \cite{10} the analysis of the UV conformal weights (\ref{DELTA}-\ref{jpm}) in the twistless 
case  ($\alpha=\omega=0$) yields that the operator content of the UV limit of the theory with ordinary periodic 
boundary conditions is that of the modular invariant $c=3/2$ CFT with partition function given by (\ref{ZR}). 
The topological charge $S$ of the model can be identified with the winding number $m$ of (\ref{ZR}), 
and $\delta_y$ is the parameter of the NLIE which distinguishes the Neveu-Schwartz (NS) and Ramond (R) sectors 
of the theory, namely $\delta_y=0$ corresponds to the Neveu-Schwartz sector, and $\delta_y=1$ describes the 
Ramond sector of the model.

 \section{ Twisted NLIE and $N=1$ superminimal models perturbed by $\Phi_{13}$}

 Based on earlier results \cite{14a,14b,15,16,6} it is expected that the NLIE (\ref{nlie1}) at appropriate 
values of
the twist angle $\alpha$ can describe the finite volume spectrum of the supersymmetry preserving perturbation
$\Phi_{13}$ of the $N=1$ superminimal models. In this section we confirm this expectation.
 
 The $N=1$ superminimal models ${\cal SM}(r,s)$ can be characterized by two positive integers $r$ and $s$ such 
that $r$ and $\frac{s-r}{2}$ are coprime integers. Their central charge is given by
\begin{equation} \label{cmin}
c(r,s)=\frac32 \left(1-2\frac{(s-r)^2}{rs}  \right).
\end{equation}
The highest weights of primary fields are also characterized by two integers, $k$ and $l$:
\begin{equation} \label{Dkl}
\Delta(k,l)=\frac{(r l-s k)^2-(r-s)^2}{8rs}+\frac{1-(-1)^{k+l}}{32}, \qquad
1 \leq k \leq r-1, \quad 1 \leq l \leq s-1.
\end{equation}
The sum $k+l$ is even in the Neveu-Schwartz sector and odd in the Ramond sector. The special choice $s=r+2$
corresponds to the series of unitary superminimal models. 
 
 The parameter $\alpha$ of the NLIE is related to the twist $\omega$ of the underlying Bethe Ansatz solvable
 lattice model as:
 \begin{equation} \label{formula1}
\alpha=\omega \frac{p+2}{p}+\chi_{\infty}(N_--N_+).
\end{equation}
 The twisted lattice Bethe equations (see section 3 of ref. \cite{10}) are invariant 
 under the change $\omega \rightarrow \omega+\pi.$
 This discrete symmetry of the integrable lattice model is reflected by our NLIE in such a way that when
 we replace $\omega$ by $\omega+\pi$, according to (\ref{formula1}) the value of $\alpha$ changes by $2 \pi$
 times an integer, but shifting $\alpha$ by $2 \pi$ is an invariance of the NLIE (\ref{nlie1}), with 
appropriate redefinition of the auxiliary functions  ($\log b,\log y,... \mbox{etc.}$) and the Bethe quantum 
numbers. 

 From now on we restrict ourselves to the case of neutral (i.e. $S=0$) states. Then the superminimal models
 ${\cal SM}(r,s)$ perturbed by $\Phi_{13}$ can be obtained by an RSOS restriction of the SSG 
model \cite{20}. This RSOS restriction is similar to the one in the ordinary sine-Gordon case \cite{21,22}.
To get access to the perturbed superminimal models we should restrict ourselves to the $S=0$ neutral charge
sector of the theory in such a way that the parameter $p$ of the NLIE must be
\begin{equation} \label{pmin}
p=\frac{2r}{s-r}
\end{equation}
 and the twist must take the values:
 \begin{equation} \label{omegak}
\omega=k \frac{\pi}{p+2} \quad \mbox{mod} \quad \pi, \qquad k \in {\mathbb Z}.
\end{equation}
All of these values of the twist $\omega$ is necessary to get all the states of the corresponding RSOS model.
On the other hand not all these twist values correspond to inequivalent values of $\alpha$ and so to
different physical states. This is a consequence of the RSOS truncation.  
 
 Following from (\ref{formula1}), (\ref{pmin}) and (\ref{omegak}) the values of $\alpha$ corresponding to
 the necessary RSOS restriction are as follows:
 \begin{equation} \label{alfak}
\alpha_k=k \frac{\pi t}{r}, \qquad k=1,2,\dots,r-1, 
\end{equation}
where we introduced the notation
$$ t=\frac{s-r}{2}.$$ 
The $t=1$ special choice corresponds to the case of the unitary superminimal models. 

Restricting ourselves to the $S=0$ neutral sector, inserting 
(\ref{alfak}), (\ref{pmin}) and (\ref{comprad}) into the general formulae of UV
conformal weights (\ref{DELTA}-\ref{jpm}), and correcting the final formula by a trivial contribution coming 
from the difference of the central charges of the $c=3/2$ and the superminimal CFTs, 
one gets the UV conformal weights as follows:
\begin{equation} \label{uvmin}
\tilde{\Delta}^{\pm}(n_{\pm},k)=\frac{\delta_y}{16}+
\frac{ ((2 n_\pm+k)r-k s)^2-(r-s)^2 }{8rs}+N_\pm+J_{\pm}.
\end{equation}
Exploiting (\ref{f1ns}) and (\ref{f1r}) one can see that this formula agrees with (\ref{Dkl}) if one identifies
the integer $l$ of (\ref{Dkl}) with $2 n_\pm +k$ in (\ref{uvmin}). 
 So as not to overflow the Kac table of the superminimal models one should impose the constraint:
\begin{equation} \label{contraint}
1\leq 2 n_\pm+k \leq s-1, \qquad k=1,2,\dots,r-1.
\end{equation}
Furthermore locality requirements (i.e. $\Delta^+-\Delta^-$ must be integer)
force us to accept only those solutions of the plateau equations where $n_+=n_-.$  
 Restricting our attention to states satisfying the previous two constraints we get the UV conformal weights
 of the ${\cal SM}(r,s)$ superminimal models. This fact suggests us that the twisted NLIE at the values
 of the parameter $\alpha$ given by (\ref{alfak}) describes the finite size effects of the integrable
 perturbations of the ${\cal SM}(r,s)$ models. 
 
 To close this section we remark that the parameter $\alpha$ does not occur in the equation of 
 $\log \tilde{a}(\theta)$ in the attractive regime. Since $\log \tilde{a}(\theta)$ is the only 
 counting function which determines the breather-breather S-matrices, the infrared asymptotics of the
 breather states does not depend on $\alpha$ and so the S-matrices involving only
 breathers are also unchanged. 
 As a consequence in accordance with \cite{20}, the NLIE yields that in the perturbations of non-unitary 
superminimal models, the breather-breather S-matrices can be calculated simply
 by inserting the appropriate value of $p$ from (\ref{pmin}) into the formula of the breather S-matrices of 
appendix A. For example in the second section we demonstrated that the $B_1-B_1$ S-matrix can be reproduced
by the NLIE. According to (\ref{pmin}) the $p=2/3$ choice corresponds to the supersymmetric Lee-Yang model, 
so substituting $p=2/3$ into the $B_1-B_1$ S-matrix, we can reproduce the well known S-matrix \cite{20} of the 
supersymmetric Lee-Yang model.

By this simple example we demonstrated that not only the UV limit of the perturbed superminimal models are
reproduced correctly, but their S-matrix as well.
Here we remark that in the unitary perturbed superminimal models, the actual value of the twist-like parameter
 $\alpha$ plays an important role in the determination of the S-matrix. In this case we have only kink 
S-matrices  and after lengthy calculations it can be shown that the twisted NLIE reproduces
 the kink-kink S-matrices \cite{23} of the perturbed unitary superminimal models.

 \section{Some examples of breather states}
 In this section we consider some of the simplest $B_1$ breather states and with the help of the NLIE 
 we determine their UV conformal weights. In the attractive regime the simplest excitations 
of the  supersymmetric sine-Gordon model are the single $B_1$ breather states at rest. 
Due to the zero momentum condition and the left-right symmetry of the counting functions,
the single self-conjugate effective root characterizing the excitation is located exactly on the 
imaginary axis, i.e. $w_{sc}=i \frac{\pi}{2}(p+1).$ Thus no quantization condition is necessary to be imposed 
on the self-conjugate effective root. 

 First let us consider the static $B_1$ breather state quantized with $\delta_y=\delta_y=0.$
Calculating the UV conformal dimensions along the lines of section 4 we obtain the conformal dimensions:
$$ \Delta^{\pm}=\frac{1}{R^2},$$ 
so the UV limit of this state is the linear combination 
$|1-\rangle=\frac{1}{\sqrt{2}}(V_{1,0}^{(0,0)}(0,0)-V_{-1,0}^{(0,0)}(0,0)|0\rangle$ 
of the vertex operators $V_{\pm 1,0}^{(0,0)}$ of the modular invariant $c=3/2$ CFT (\ref{ZR}). 

 Using conformal perturbation theory (PCFT) the leading order
correction to this state has been calculated in \cite{18}, and it reads as:
\begin{equation} \label{cpt1}
\delta \varepsilon_{|1-\rangle}(\ell)= \frac{6L}{\pi} E_{|1-\rangle}(\ell)+\left(\frac32-12(\Delta^+ 
+\Delta^-)\right)=
-C_2^{|1-\rangle} \, \ell^{2y}+O(\ell^{4y}),
\end{equation} 
where $y=1-\frac{1}{R^2}$ and the leading order coefficient takes the form:
\begin{equation} \label{cpt2}
C_2^{|1-\rangle}=\alpha \, \tilde{I}_{|1-\rangle},
\end{equation}
where
\begin{equation} \label{cpt3}
\alpha=\frac{3}{2}\cdot8^{{2}/{R^{2}}}\frac{1}{\gamma^{2}\left(\frac{1}{2}-
\frac{1}{2R^{2}}\right)}\left(\frac{1}{R^{2}-1}\right)^{2-\frac{2}{R^{2}}}, \qquad
\gamma(x)=\frac{\Gamma(x)}{\Gamma(1-x)},
\end{equation}
and
\begin{equation} \label{cpt4}
\tilde{I}_{|1-\rangle}=\gamma\left(\frac{1}{2}-\frac{1}{2R^{2}}\right)\left[\gamma\left(-\frac{1}{R^{2}}
\right)
\gamma\left(\frac{1}{2}+\frac{3}{2R^{2}}\right)-\frac{1}{2}\gamma\left(\frac{1}{2}-\frac{1}{2R^{2}}\right)
\gamma\left(\frac{1}{R^{2}}\right)\right].
\end{equation}
We checked numerically our NLIE against PCFT at a lot of values of the compactification radius $R$,
 and in every case we experienced very good agreement. To illustrate the agreement between NLIE and
PCFT, the numerical comparison of $\delta \varepsilon_{|1-\rangle}(\ell)$
coming from NLIE and leading order PCFT at the specific $R^{2}=5$ point can be found in table 1. 
One can see that the two sets of data approach to one another as the volume tends to zero.

\begin{table}
\begin{center}\begin{tabular}{|c|c|c|c|}
\hline 
 $\ell$&
 $\delta \varepsilon_{|1-\rangle}(\ell) \quad\mbox{(NLIE)}$&
 $\delta \varepsilon_{|1-\rangle}(\ell)  \quad\mbox{(PCFT)}$&
 $\ell^{4y}$\tabularnewline
\hline
0.1 & 0.028726870911825 & 0.02873952098276545 & $6.3\cdot10^{-4}$\tabularnewline
\hline
0.05 & 0.009479129734511 & 0.00948050632214565 & $6.86 \cdot 10^{-5}$\tabularnewline
\hline
0.01 & 0.000721896146209 & 0.00072190412804693 & $3.98\cdot10^{-7}$\tabularnewline
\hline
0.005 & 0.000238138683370 & 0.00023813955194438 & $4.33\cdot10^{-8}$\tabularnewline
\hline
0.001 & 0.000018133406805 & 0.00001813341184091 & $2.51 \cdot 10^{-10}$\tabularnewline
\hline
0.0005 & 0.000005981794545 & 0.00000598179509334 & $2.73\cdot10^{-11}$\tabularnewline
\hline
0.0001 & 0.00000045549071 &  0.00000045549071160178 & $1.58\cdot10^{-13}$ \tabularnewline
\hline
\end{tabular}\end{center}

\caption{{\footnotesize Numerical comparison of PCFT with NLIE for the state
$|1-\rangle$ at $R^{2}=5$. }}
\end{table}

Another important example of the one $B_1$ breather states is the static one $B_1$ breather state in the
Ramond sector. This state is quantized by $\delta_y=1,$ $\delta_b=0.$ The UV conformal weight corresponding to
this state is 
\begin{equation} \label{wr1}
 \Delta^{\pm}=\frac{1}{16}+\frac{1}{8 R^2},
\end{equation}
which turns out to be the linear combination of the vertex operators $R_{\pm\frac12,0}(0,0).$

To close this section we also calculate the UV conformal weights of some of the two $B_1$ breather states 
with lowest energy and zero momentum. These states contain two self-conjugate effective roots. 
One of them is a left mover and the other is a right mover in the UV limit. 
According to section 3 these states may contain type I holes as well. So in this case not only the quantum 
numbers $\delta_b$ and $\delta_y$ characterize the lowest lying state but the number of type I holes as well.   
Performing the calculation of the UV conformal weights of the lowest lying zero momentum two $B_1$ states
one gets that the state having quantum numbers $\delta_y=\delta_b=N_1=0$ corresponds to the symmetric
first level descendent of the vacuum with weights:
$$  \Delta^{\pm}=1.$$
The state quantized with quantum numbers $\delta_y=\delta_b=0$ and $N_1=2$ has UV weights
$$ \Delta^{\pm}=1/2,$$
corresponding to the state $\bar{\psi}\psi(0,0)|0\rangle$.

In the Ramond sector (i.e. $\delta_y=1$) the two $B_1$ state, which has quantum numbers $\delta_b=1$ and 
$N_1=0$ in the infrared limit, will have quantum numbers $N_1=2,$ $N_1^\pm=1$ in the UV limit due to the fact
that the counting function $i \, \log a(\theta)$ will have a negative slope part creating new type I holes.
The calculation of the UV conformal weights of this state yields:
$$\Delta^{\pm}=\frac{1}{16}+\frac{1}{8 R^2},$$
which turns out to be the same as that of the one $B_1$ state at rest of the Ramond sector (\ref{wr1}), so this
state must originate from the other linearly independent combination of the vertex operators
$R_{\pm 1/2,0}(0,0)$ in the UV.

\section{Numerical test of the twisted NLIE}
The most obvious test of the twisted NLIEs for the perturbed superminimal models would be the numerical
comparison against data coming from truncated conformal space approach (TCSA). Unfortunately there is no
available TCSA data in the literature for this class of models. This is why we choose another way to
test the twisted NLIEs. It is well known that the $c=7/10$ unitary CFT is present in the unitary series
of superminimal and Virasoro minimal models as well. In the superminimal language it corresponds to 
${\cal SM}(3,5)$ whilst in the minimal CFT language it corresponds to ${\cal M}(4,5)$ and certainly their
integrability preserving perturbations are also equivalent. In ref. \cite{6} the NLIE description
of the finite size effects in minimal models perturbed by $\Phi_{13}$ was given introducing an appropriate
twist parameter into the NLIE of the sine-Gordon model. For later convenience we will refer to these equations 
as twisted Destri-de Vega (DDV) equations. So the finite size spectrum of the ${\cal M}(4,5)+\Phi_{13}$ model
can be described by two different sets of NLIEs. First considering the model as a perturbed superminimal 
model the set of NLIEs is given by the equations (\ref{nlie1}-\ref{eq:t1}) with $p=3$ and twist values 
$\alpha_k=\frac{k\pi}{3}$. Then formulating the model as a perturbed minimal 
model the finite size effects are governed by the twisted DDV equations of ref. \cite{6} with $p_{DDV}=4$ and 
twist values $\alpha_{DDV}^{(k)}=\frac{k\pi}{4}$. 

Hereinafter we will compare numerically the two types of NLIEs for the ground states of the 
${\cal SM}(3,5)+\Phi_{13}$ model. It is well known that the ${\cal SM}(3,5)+\Phi_{13}$ model has 3 ground 
states which are degenerate in infinite volume, but they split as the volume is decreased. In the UV 
the 3 ground states of the model correspond to primaries with conformal weights 
$\Delta^{\pm} \in \{ 0,\frac{3}{80},\frac{1}{10}\}$. In the language of the twisted DDV equations
they correspond to twist values: $\alpha_{DDV} \in \{\frac{\pi}{4},\frac{\pi}{2},\frac{3\pi}{4} \}$ 
respectively. In the language of our twisted NLIEs (\ref{nlie1}-\ref{eq:t1}) 
the ground states with UV conformal weights
$0$ and $\frac{1}{10}$ can be found in the NS sector of the model $(\delta_y=0)$ and they correspond to
twist values: $\alpha=\frac{\pi}{3}$ and $\alpha=\frac{2\pi}{3}$ respectively. The ground state with 
$\Delta^{\pm}=\frac{1}{10}$ is in the R sector $(\delta_y=1)$ and the corresponding twist value is 
$\alpha=\frac{2\pi}{3}$. The numerical comparison of the two types of NLIEs for the 3 ground states of the
model can be found in tables 2,3 and 4. The high accuracy agreement between the two sets of numerical data
reassures the correctness of our twisted NLIEs for the ${\cal SM}(r,s)+\Phi_{13}$ models.

\begin{table}
\begin{center}\begin{tabular}{|c|c|c|}
\hline 
 $\ell$&
 $E_1 \quad\mbox{(NLIE)}$&
 $E_1  \quad\mbox{(DDV)}$\tabularnewline
\hline
2 & $-0.0597271103$ & $-0.0597271103$ \tabularnewline
\hline
1 & $-0.2399086782$ & $-0.2399086782$  \tabularnewline
\hline
0.5 & $-0.6282004195$ & $-0.6282004195$  \tabularnewline
\hline
0.1 & $-3.6166077208$ & $-3.6166077208$ \tabularnewline
\hline
0.05 & $-7.2973738367$ & $-7.2973738367$ \tabularnewline
\hline
0.01 & $-36.6389894487$ & $-36.6389894489$ \tabularnewline
\hline
\end{tabular}\end{center}

\caption{{\footnotesize The first ground state of the superminimal model ${\cal SM}(3,5)$ perturbed
by $\Phi_{13}$ corresponding to $\Delta^{\pm}=0$.}}
\end{table}

\begin{table}
\begin{center}\begin{tabular}{|c|c|c|}
\hline 
 $\ell$&
 $E_2 \quad\mbox{(NLIE)}$&
 $E_2  \quad\mbox{(DDV)}$\tabularnewline
\hline
2 & $0.0007298114$ & $0.0007298114$ \tabularnewline
\hline
1 & $0.0109172291$ & $0.0109172291$  \tabularnewline
\hline
0.5 & $0.0613220318$ & $0.0613220318$  \tabularnewline
\hline
0.1 & $0.7610380973$ & $0.7610380973$ \tabularnewline
\hline
0.05 & $1.7468225711$ & $1.7468225711$ \tabularnewline
\hline
0.01 & $9.9655028929$ & $ 9.9655028911$ \tabularnewline
\hline
\end{tabular}\end{center}

\caption{{\footnotesize The second ground state of the superminimal model ${\cal SM}(3,5)$ perturbed
by $\Phi_{13}$ corresponding to $\Delta^{\pm}=\frac{3}{80}$.}}
\end{table}

\begin{table}
\begin{center}\begin{tabular}{|c|c|c|}
\hline 
 $\ell$&
 $E_3 \quad\mbox{(NLIE)}$&
 $E_3  \quad\mbox{(DDV)}$\tabularnewline
\hline
2.5 & $0.0344866621$ & $0.0344866621$ \tabularnewline
\hline
2 & $0.0668483893$ & $0.0668483893$ \tabularnewline
\hline
1.5 & $0.1380515377$ & $0.1380515377$ \tabularnewline
\hline
1 & $0.3217601530$ & $0.3217601530$  \tabularnewline
\hline
0.8 & $0.4812047615$ & $0.4812047615$  \tabularnewline
\hline
0.5 & $1.0154869429$ & $1.0154869429$  \tabularnewline
\hline
\end{tabular}\end{center}

\caption{{\footnotesize The third ground state of the superminimal model ${\cal SM}(3,5)$ perturbed
by $\Phi_{13}$ corresponding to $\Delta^{\pm}=\frac{1}{10}$. For $\ell \lesssim 0.5$ special objects appear in 
the
NLIEs obstacling the convergence of the usual numerical iteration.  }}
\end{table}

\section{Summary and perspectives}

In this paper we presented and analysed the NLIE governing the finite size effects in both the
attractive and repulsive regimes of the SSG. 
We also demonstrated that the twisted version of the NLIE at certain particular values 
of the twist parameter describes the finite size effects in the $N=1$ superminimal models 
perturbed by $\Phi_{13}.$    

As a starting point, we extended our previous studies \cite{10} of the SSG 
to the attractive regime in order to describe breathers. We showed that in the infrared limit 
the NLIE successfully reproduces the scattering  amplitudes involving breathers and
we gave examples of comparing numerical results from the NLIE to those coming from conformal
perturbation theory. 
 
We then proceeded to the case of perturbed superminimal models. Following the idea of the introduction of 
a twist parameter \cite{6,14a,14b,15,16}, 
we have shown that in the UV limit the twisted NLIE at certain specific 
values of the twist angle gives conformal weights that are consistent with the spectrum of $N=1$ superminimal 
CFTs. It turned out that similarly to the case of perturbed Virasoro minimal CFTs \cite{6} 
to describe all possible states of the limiting superminimal CFT it is not enough to choose a single value of 
the twist parameter, but a range of values is required (\ref{alfak}). In the IR limit
we demonstrated that the twisted NLIE reproduces correctly the bootstrap S-matrices of perturbed superminimal 
models. In the special case of 
${\cal SM}(3,5)+\Phi_{13}$ model we checked numerically our NLIEs against the twisted DDV equations of ref.
\cite{6} and we found perfect agreement.
We think that the evidences presented in this paper strongly supports the validity of the NLIE description
of perturbed superminimal models.
 
 The SSG is just the first in a whole series fractional 
supersymmetric sine-Gordon models \cite{25Fr} that can be seen as perturbations of the conformal
models introduced in \cite{17Fr}. The full series of coupled NLIEs generalizing the ones presented here
has been proposed, for the vacuum by C. Dunning \cite{27Fr}. It would be nice to extend Dunning's 
conjecture to excited states as well. To achieve this one has to start from higher spin vertex models
and along the lines of \cite{9} and extend the method described in \cite{8} 
 for excited states. Having the NLIEs for the excited states two applications can be of 
importance: the quantum group restriction leading to $SU(2)$-coset theories \cite{30Fr} perturbed by
$\Phi_{13}^{(1)}$ and the large spin limit where this series should make contact with the
$N=2$ supersymmetric sine-Gordon model, whose finite size effects are of principal
importance in the formulation  of superstrings propagating in pp-wave background \cite{31Fr}.

Recently in \cite{19} the vacuum state of the SSG with Dirichlet boundary 
conditions has been investigated by means of the NLIE technique. The extension of these NLIEs to
excited states and to more general boundary conditions would be of interest. 

We hope to return to these issues in the future.

 \vspace{1cm}
{\tt Acknowledgements}

\noindent 
This investigation was supported by the Hungarian National Science Fund OTKA (under T049495).

\appendix      

\renewcommand{\thesection}{Appendix~A.}
\section{}
\renewcommand{\thesection}{A}

This appendix is devoted to recall the breather-breather S-matrices of the supersymmetric 
sine-Gordon model \cite{20}. The \(n\)th breather has mass
\begin{equation}
m_n=2{\cal M}\sin\left(\frac{n p \pi}{2}\right)\,.
\end{equation}
The breather 2-body S-matrix can be written in the form \cite{20}
\begin{equation}
S_\text{SSG}^{(n,m)}(\theta)=S_\text{SG}^{(n,m)}(\theta)S_\text{SUSY}^{(n,m)}(\theta)\,
\end{equation}
where
\begin{multline}
S_\text{SG}^{(n,m)}(\theta)=\frac{\sinh(\theta)+i\sin(\frac{n+m}{2}p \pi)}{\sinh(\theta)-i\sin(\frac{n+m}
{2}p \pi)}
  \frac{\sinh(\theta)+i\sin(\frac{n-m}{2}p \pi)}{\sinh(\theta)-i\sin(\frac{n-m}{2}p \pi)}\times\\
\prod_{k=1}^{m-1}\frac{\sin^2(\frac{n-m-2k}{4}p \pi+i\frac\theta2)}{\sin^2(\frac{n-m-2k}{4}p \pi
-i\frac\theta2)}\frac{\cos^2(\frac{n+m-2k}{4} p \pi+i\frac\theta2)}{\cos^2(\frac{n+m-2k}{4} p \pi
-i\frac\theta2)}
\end{multline}
is the sine-Gordon S-matrix and the supersymmetric factor is
\begin{equation}
S_\text{SUSY}^{(n,m)}(\theta)=M^{(n,m)}(\theta)G^{(n,m)}(\theta)
\end{equation}
with $M^{(n,m)}(\theta)$ being a $4\times 4$ matrix of the form
\begin{multline}
M^{(n,m)}(\theta)=\\
\begin{pmatrix}
1+i\,\frac{\sin(\frac{n p \pi}{2})+\sin(\frac{m p \pi}{2})}{\sinh(\theta)}&0&0
&\frac{\sqrt{\sin(\frac{n p \pi}{2})\sin(\frac{m p \pi}{2})}}{\cosh(\frac\theta2)}\\
0&1-i\,\frac{\sin(\frac{n p \pi}{2})-\sin(\frac{m p \pi}{2})}{\sinh(\theta)}&i\,
\frac{\sqrt{\sin(\frac{n p \pi}{2})\sin(\frac{m p \pi}{2})}}{\sinh(\frac\theta2)}&0\\
0&i\,\frac{\sqrt{\sin(\frac{n p \pi}{2})\sin(\frac{m p \pi}{2})}}{\sinh(\frac\theta2)}
&1+i\,\frac{\sin(\frac{n p \pi}{2})-\sin(\frac{m p \pi}{2})}{\sinh(\theta)}&0\\
\frac{\sqrt{\sin(\frac{n p \pi}{2})\sin(\frac{m p \pi}{2})}}{\cosh(\frac\theta2)}&0&0&-1+i\,
\frac{\sin(\frac{n p \pi}{2})+\sin(\frac{m p \pi}{2})}{\sinh(\theta)}\\
\end{pmatrix}
\end{multline}
and $G^{(n,m)}(\theta)$ is given by
\begin{gather}
G^{(n,m)}(\theta)=\frac{g(\frac{n+m}{4}p|\theta)g(\frac12-\frac{n-m}{4}p|\theta)}{g(\frac12|\theta)}\,,\\
g(x|\theta)=\frac{\sinh(\frac\theta2)}{\sinh(\frac\theta2)+i\sin(x\pi)}
  \exp\left\{\int_0^\infty\frac{d q}{q}
  \frac{\sinh(x \pi q)\sinh((1-x)\pi q)}{\cosh^2(\frac{\pi q}2)\cosh(\pi q)}
  \sinh\left(i q \theta \right)\right\}\,.
\end{gather}

Now we restrict our attention to the $n=m=1$ case. In this sector there are 4 eigenvalues of the 2-body
breather S-matrix $S^{(1,1)}(\theta)$. The first pair of eigenvalues is of the form:
\begin{equation} \label{ev1pm}
\Lambda^{(1)}_{\pm}(\theta)=\lambda^{(1)}_{\pm}(\theta) G^{(1,1)}(\theta) S_{SG}^{(1,1)}(\theta),
\end{equation} 
where
$$\lambda^{(1)}_{\pm}(\theta)=1\pm i \frac{\sin\left(\frac{p \pi}{2} \right)}{\sinh\left( \frac\theta2 
\right)}. $$

The second pair of eigenvalues takes the form:
\begin{equation} \label{ev2pm}
\Lambda^{(2)}_{\pm}(\theta)=\lambda^{(2)}_{\pm}(\theta) G^{(1,1)}(\theta) S_{SG}^{(1,1)}(\theta),
\end{equation} 
where
$$\lambda^{(2)}_{\pm}(\theta)=2 i \frac{\sin\left(\frac{p \pi}{2} \right)}{\sinh\left(\theta 
\right)} \pm \sqrt{\frac{\sin^2\left(\frac{p \pi}{2} \right)}{\cosh^2\left( \frac\theta2 
\right)}+1}. $$

\renewcommand{\thesection}{Appendix~B.}
\section{}
\renewcommand{\thesection}{B}

In this appendix we present the result of the integral (\ref{Ih0}) in case of $\delta_y=0.$
Inserting (\ref{y12}) into (\ref{Ih0}) and calculating the integral finally one gets:
\begin{equation} \label{ih00}
I_h^{(0)}(w_{sc})|_{w_{sc}=\theta+i \frac{\pi(p+1)}{2}}=\hat{I}_f(\theta-\theta_{12})-F_0(\theta-\theta_1)-
F_0(\theta-\theta_2),
\end{equation}
where
\begin{equation} \label{F0}
F_0(\theta)=-\ln \frac{\cosh\left( \frac{\theta-i p \pi}{4} \right)}{\cosh\left( \frac{\theta+i p \pi}{4} 
\right)}+i \, \chi_G(\theta).
\end{equation}
and
\begin{equation} \label{ifhat}
\hat{I}_f(\theta)=G(e^{\theta+i p \pi/2})-G(e^{\theta-i p \pi/2})
+\frac{(A_+^4-1)W({e^\theta}/{A_+})}{A_+^2(A_+^2-A_-^2)}
-\frac{(A_-^4-1)W({e^\theta}/{A_-})}{A_-^2(A_+^2-A_-^2)}
\end{equation}
The functions and constants occurring in (\ref{ifhat}) are as follows:
\begin{equation} \label{}
W(z)=\mbox{arctanh}(e^{ip\pi/2}z)-\mbox{arctanh}(e^{-ip\pi/2}z),
\end{equation}
\begin{equation} \label{D}
D=\cosh^2\left(\frac{\theta_1-\theta_2}{2}\right)+\sin^2\left(\frac{p \pi}{2} \right),
\end{equation}
\begin{equation} \label{}
A^2=4D-2=2 \cosh(\theta_1-\theta_2)+4 \sin^2\left(\frac{p \pi}{2} \right),
\end{equation}
\begin{equation} \label{A+-}
A_{\pm}=\sqrt{D}\pm \sqrt{D-1},
\end{equation}
\begin{equation} \label{}
G(z)=-\ln z+ \frac12 \ln (z^4-A^2 z^2+1).
\end{equation}

\renewcommand{\thesection}{Appendix~C.}
\section{}
\renewcommand{\thesection}{C}

In section \ref{sec:UV}, in the calculation of the UV conformal weights,
the following dilogarithmic sum must be calculated: \begin{equation}
S_{0}(\rho)=2\left\{ L_{+}\left[e^{3i\rho}
\,2\,\cos(\rho)\right]+L_{+}\left[e^{-3i\rho}\,2\,\cos(\rho)\right]+L_{+}\left[
\frac{\sin\left(3\rho\right)}{\sin\left(\rho\right)}\right]\right\} ,\label{S0}\end{equation}
 where \begin{equation}
L_{+}(x)=\frac{1}{2}\int\limits _{0}^{x}\, dy\,\left\{ \frac{\ln(1+y)}{y}-
\frac{\ln y}{1+y}\right\}.\end{equation}
 The function $S_{0}(\rho)$ has the following simple properties:
\begin{equation}
S_{0}(\rho)=S_{0}(\rho+\pi),\qquad S_{0}(\rho)=S_{0}(-\rho).\end{equation}
 After some algebra one can prove that \begin{equation}
S_{0}(\rho)=\frac{2\pi^{2}}{3}+\hat{N_{\rho}}
\left\{ 2\pi\left|\rho-\pi\left[\frac{\rho}{\pi}\right]-\frac{\pi}{2}\right|-\pi^{2}\right\} 
-i\,\hat{N_{\rho}}\pi\,\ln\left(4\,\cos^{2}\rho\right),\label{RS0}\end{equation}
 where \begin{equation}
\hat{N}_{\rho}=N_{\rho}\,\,\mbox{mod}\,\,2\qquad 
N_{\rho}=\left[3\frac{\rho}{\pi}\right]-\left[\frac{\rho}{\pi}\right],\end{equation}
 and we made the choice of $\ln(-1)=i\pi$.

\renewcommand{\thesection}{Appendix~D.}
\section{}
\renewcommand{\thesection}{D}
   
In this appendix the most important properties of the function $\chi_K(\theta)$ will be clarified. 
 The function $\chi_K(\theta)$ is defined as the odd primitive of the kernel $2 \pi \, K(\theta)$ and 
given by the formula:
 \begin{equation} \label{}
\chi_K(\theta)=-i \, \ln \frac{\sinh\left( i \frac{\pi}{4}-\frac{\theta}{2} \right)}
{\sinh\left( i \frac{\pi}{4}+\frac{\theta}{2} \right)}, \quad
\chi_K(\theta)=-\chi_K(-\theta) \quad  \forall \theta \in {\mathbb C}.
\end{equation}
The branch cuts are chosen to run parallel to the real axis so that $\chi_K(\theta)$ be an odd real 
analytic function on the entire complex 
plane and continuous along the real axis. In this case $\chi_K(\theta)$ is not periodic  anymore with 
respect to $2 \pi i$. It is periodic only modulo $2 \pi$, i.e. the following identity holds:
$$\chi_K(\theta+2\pi i)=\chi_K(\theta)-2\pi.$$ It follows that the distance between the consecutive cuts
is $2 \pi i$ and the jump of $\chi_K(\theta)$ is equal to $-2\pi$ at each branch cut crossed from 
below to up. 
The choice of branch cuts is depicted in figure 1.  
\begin{figure}[htb]
\begin{flushleft}
\hskip 15mm
\leavevmode
\epsfxsize=160mm
\epsfbox{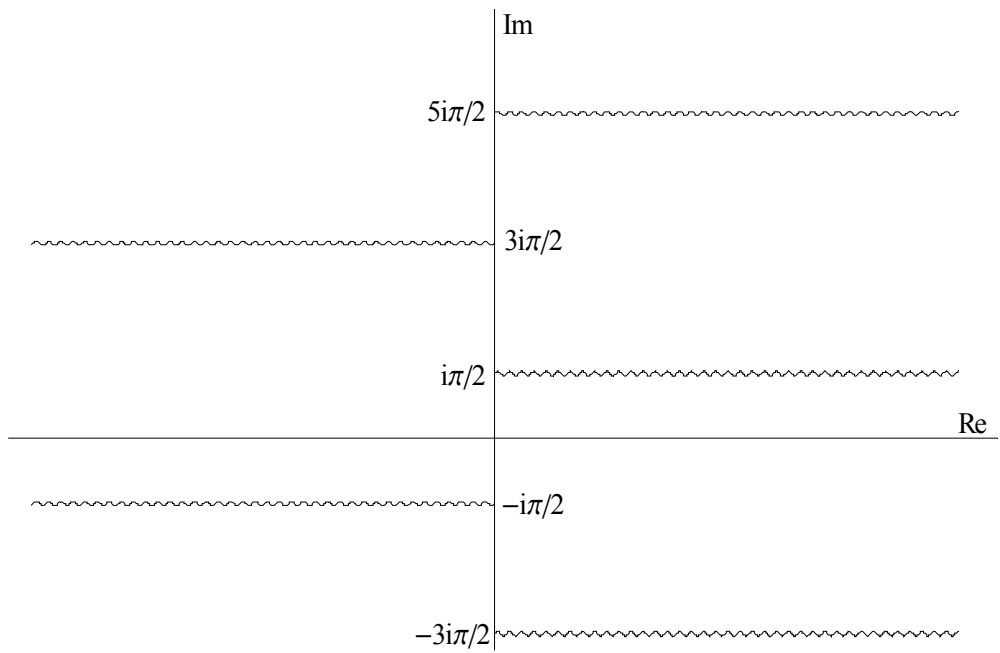}
\end{flushleft}
\caption{{\footnotesize
Locations of the branch cuts of $\chi_K(\theta).$
}}
\label{6}
\end{figure}

\end{document}